# Matrix-Driven Quartic Overhauser (QOVR) Surface Structural Framework: Continuity Limitations, Computer Graphics Algorithms, and Software Implementation

*A Preprint.*

Hakan Üstünel
(0000-0001-9903-593X)

*Department of Software Engineering, Engineering Faculty, Kırklareli University, 39100, Kırklareli, Türkiye; hakanustunel@klu.edu.tr, hakanustunel@hotmail.com*

**Abstract:** This study introduces the spatial and analytical construction of the Quartic Overhauser (QOVR) surface generation framework designed to address boundary alignment and localized shape modification constraints. This framework implements a variable-parameter fourth-degree novel matrix formulation to achieve parameter tracking across orthogonal coordinate axes. The analytical pipeline integrates directional spline blending functions with symmetric spatial control matrices, ensuring that internal knot vector variations allow localized surface adjustments while maintaining the position convergence of global edge boundaries. Computational verification confirms that while the current formulation satisfies explicit $C^0$ positional closure and $C^1$ tangent continuity conditions across the internal and boundary interfaces without triggering global curvature propagation, with edge joint separation, or wave-like artefacts, $C^2$ curvature continuity is not maintained. As demonstrated by three-dimensional mesh models and colormap visualizations, the spatial intensity fields condense strictly within the immediate neighborhood of the modified element, verifying that the displacement effect decreases progressively as the distance from the perturbed control points increases. This explicit decoupling aligns with structural symmetry and boundary alignment across adjacent geometric patches, satisfying manufacturing and reverse engineering sealing requirements.

## 1. Introduction

Parametric curve manipulation underpins continuous spatial network synthesis in computer-aided geometric design (CAGD) and computer graphics. Conventional B-spline and Bézier representations employ blending functions controlled by static internal knots under explicit control point configurations, ensuring global continuity but requiring coordinated inputs for localized shape adaptation. Consequently, internal control point per-mutations affect the curvature propagation or edge coordinate displacements, disrupting global boundary alignment. This constraint represents a key consideration in procedural content generation (PCG) systems and industrial CAD/CAM workflows requiring seamlessly anchored geometric patches to maintain defined sealing criteria [1,2].

To address these boundary alignment and localized modification constraints, this study introduces a matrix-driven Quartic Overhauser (QOVR) surface framework. Derived from the variable-parameter fourth-degree spline architecture [3], this model tracks parametric adjustments across orthogonal coordinate axes. Integrating directional spline blending functions with spatial control matrices enables localized surface optimization via internal knot vector variations while maintaining positional invariance along global edge boundaries. Computational verification confirms $C^0$ positional closure and $C^1$ tangent continuity across internal interface s [4], though $C^2$ curvature continuity is not maintained. Spatial intensity fields condense within the neighborhood of the modified element, verifying that displacement effects decrease progressively to the distance from control points. This decoupling mechanism establishes a consistent mathematical schema design and reverse engineering applications requiring independent patch optimization under defined global boundary constraints. Aligned with the numerical foundations required for sophisticated shape engineering, this work extends the previously verified QOVR spline framework [3] into a multidimensional geometric space, delivering a mathematical approach to address boundary synchronization challenges in shape and surface modeling.

The main contributions of this study are summarized as follows: (a) formulating a matrix-driven QOVR surface equation via direction-dependent parameter transformations for spatial path tracking, (b) evaluating zero-order positional closure and first-order tangent limits across interfaces while defining the analytical limitations of second-order curvature continuity under static networks, (c) implementing a demonstrative software routine with a background data dependency for framework verification, and (d) verifying algebraic surface framework deployment tracks through visualization and shape control sensitivity.

## 2. Related Work

Parametric surface modeling constructs the mathematical schema for CAGD, computer graphics, and industrial design pipelines [5]. Classical Bézier and B-spline formulations utilize polynomial blending functions to define spatial geometry under explicit control point networks, but internal control point modifications affect the curvature distribution across adjacent surface regions [6]. This propagation requires coordinated tracking during independent patch optimization workflows in PCG and CAD/CAM systems.

To address global deformation errors, non-uniform rational B-splines (NURBSs) introduce weighting parameters, yet local adjustments alter parametric distributions and require coordinated inputs to maintain boundary alignment [7]. Alternatively, Overhauser (OVR) splines enforce automatic tangent continuity via sequential overlapping quadratic curves, avoiding complex boundary equations. Parabolic blending (PB) models are utilized to estimate thickness distributions in thermoformed components [8], while generalized parabolic blending (GPB) software frameworks alter the prediction output over default configurations [9]. This trajectory established the variable-parameter fourth-degree spline architecture, expanding localized shape manipulation via object-oriented prototyping frameworks [3] regulated by classical Bernstein polynomials [10]. Historically, Overhauser elements within boundary element methods (BEMs) eliminated slope discontinuities on free surfaces during stable time stepping operations [11]. For alternative spline schemas, quartic Catmull–Rom formulations optimize internal energy via free parameters [12], parabolic blending secures slope continuity at multi-patch junctions natively [4], and piecewise cubic methods incorporate geodesic vector routing over triangular meshes to complement traditional transformations and curve definitions [5,13]. Multi-surface paradigms implement hierarchical T-meshes to enforce $C^1$ continuity and affine invariance [14], while generalized approaches employ blended $(\alpha, \lambda, s)$-Bernstein bases to modify shape control [15].

Advanced data fitting paradigms include successive over-relaxation iterative approximation (SOR-PIA) for non-uniform cubic B-splines to alter iteration steps overhead, and shell space variational techniques for curve design [6,16]. Correspondingly, variational continuous mapping models modify [1] B-spline surface developability without explicit data parametrization. To address topological boundaries, surface blending configurations employ semi-structured B-spline layers skinned from mismatched knot vectors to bridge base fields at $C^1$ or $C^2$ intervals [2].

In parallel with industrial geometric workflows, digital character representation increasingly relies on structured mesh topologies and directional symmetries [17]. Industrial mesh refinement leverages specialised routines to generate triangular surface meshes spanning multiple fields for NURBS-enhanced finite element models without model defeaturing [18]. To evaluate these networks, periodic B-spline interpolation conditions secure matrix invertibility during control point reduction over serial closed contours [19]. Domain parameterization routines utilize improved degree elevation and knot insertion algorithms to maintain parametric consistency across multi-sided fields [20], while genetic algorithms paired with least squares fitting convert partial differential equation (PDE) surface representations into industry standard NURBS networks [21].

To target the identified literature gap, this study builds upon the recently formulated variable-parameter fourth-degree spline architecture [3] to establish the QOVR surface framework. This architecture enforces structural separation between boundary constraints and internal parametric transitions, ensuring that internal knot vector variations allow localized surface adjustments while maintaining the position convergence along global edge boundaries. Computational verification confirms that while this surface formulation satisfies explicit $C^0$ positional closure and $C^1$ tangent continuity conditions across internal interfaces without triggering edge joint separation or interface tearing under variable threshold states, $C^2$ curvature continuity is not maintained. This explicit decoupling mechanism alleviates the boundary synchronization problems of traditional formulations, providing a consistent mathematical model for interactive engineering design and reverse engineering applications requiring independent patch optimization.

## 3. Geometric Construction of the QOVR Surface

This section presents the matrix-driven mathematical architecture and algebraic equations used to construct the QOVR surface network within a CAGD environment. The geometric synthesis pipeline defines the spatial coordinates by combining direction-dependent parameter transformations with symmetric matrix–tensor operations. To maintain structural closure across the geometric network, the mathematical tracking structures the directional blending layers along orthogonal coordinate axes. This framework enables shape modification along localized network lines while providing the basis for the differential boundary evaluations to confirm interface continuity.

### 3.1 Underlying QOVR Spline Equations

Before constructing the spatial surface equations, the mathematical formulation of the underlying QOVR spline formulation is presented [3]. To prevent parameter mixing and notation conflicts during the transition from a single-parameter curve to a multi-parametric surface network, the curve parameter $t$ is mapped to the surface parameter $u$ in the interval $[0.0,1.0]$. As illustrated in Figure 1, the central segments of the curve are generated through localized blending operations. Based on the configuration of internal parametric knots, the computational framework constructs parametric transformation functions.

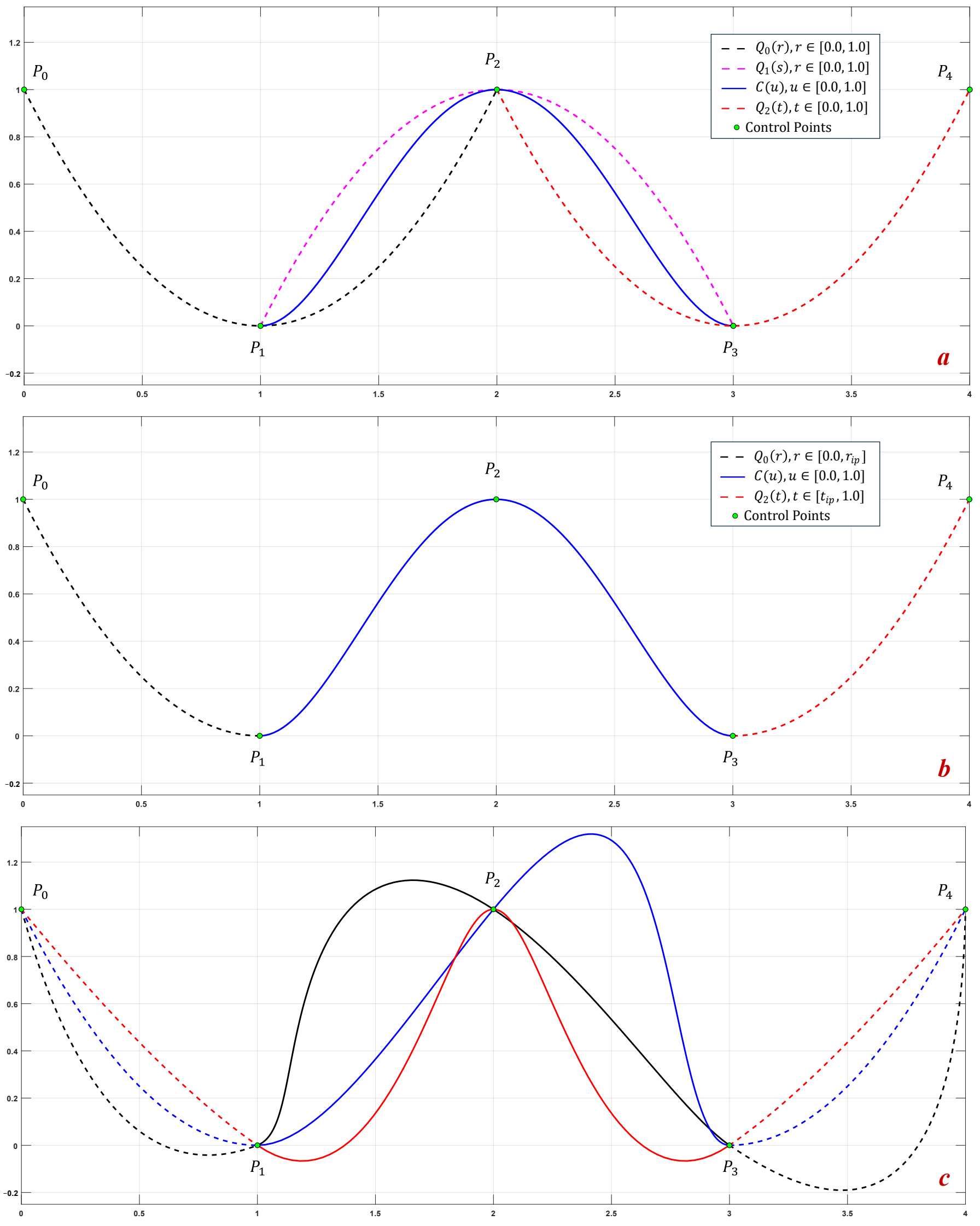

**Figure 1.** QOVR spline evaluation under parameter $u$: **(a)** independent parabolas and base blending segments, **(b)** localized internal knots with auxiliary segments, **(c)** local shape variation due to parameter modification.

The algebraic composition of the blended central segment illustrated in Figure 1 is structured through weighted linear combinations. The localized segment trajectory defined in (1) defines the interpolation boundaries by coupling the weight functions $w_i$ with the geometric outputs of the independent parabolas $Q_i$.

$$C(u) = w_0 . Q_0 + w_1 . Q_1 + w_2 . Q_2 . \quad (1)$$

To capture the geometric behavior across the intervals defined in Table 1, the mathematical formulation integrates local parameters for independent parabolas within the domains $r, s, t \in [0.0, 1.0]$. Mapping these localized variables alongside the global parameter updates the algebraic representation of the spline segment, as expressed in (2):

$$C(u) = w_0(u) * Q_0(r) + w_1(u) * Q_1(s) + w_2(u) * Q_2(t) \quad (2)$$

**Table 1.** Definitions and domains of the parameters used in the QOVR surface.

| Parameter/Symbol | Domain | Description |
|---|---|---|
| $P_{0,0} \dots P_{i,j}$ | $\mathbb{R}^d$ | Control points defining the surface. |
| $QOVR(u, v)$ | $(u, v) \in [0.0,1.0] \times [0.0,1.0]$ | Tensor product QOVR surface |
| $C(u)$ | $u \in [0.0,1.0]$ | Horizontal spline blending function. |
| $K(v)$ | $v \in [0.0,1.0]$ | Vertical spline blending function |
| $u, v$ | $[0.0, 1.0]$ | Global parametric coordinates define the surface. |
| $i, j$ | $\{0,1, \dots ,4\}$ | Parametric indices. |
| $M_{0_{C_i}}, M_{1_{C_i}}, M_{2_{C_i}}$ | Matrix space | Horizontal spline matrix kernels for the initial, central blending, and terminal stages. |
| $M_{0_{K_j}}, M_{1_{K_j}}, M_{2_{K_j}}$ | Matrix space | Vertical spline matrix kernels for the initial, central blending, and terminal stages. |
| $Q_0(r), Q_1(s), Q_2(t)$ | $\mathbb{R}^d$ | Independent parabolas defined over localized domain intervals. |
| $r, s, t$ | $[0.0, 1.0]$ | Static local parameters for independent parabolas. |
| $r_{ip_C}, s_{ip_C}, t_{ip_C}, u_{ip}$ | $(0.0, 1.0)$ | Row-wise internal parametric knots. |
| $r_{ip_K}, s_{ip_K}, t_{ip_K}, v_{ip_K}$ | $(0.0, 1.0)$ | Column-wise internal parametric knots. |
| $r(u), s(u), t(u)$ | Function mapping | Horizontal transformation functions for global to local tracking. |
| $r_u, s_u, t_u$ | Dependent on $u$ | Dynamically transformed horizontal parameters via corresponding transformation functions. |
| $r(v), s(v), t(v)$ | Function mapping | Vertical transformation functions for global to local tracking. |
| $r_v, s_v, t_v$ | Dependent on $v$ | Dynamically transformed vertical parameters via corresponding transformation functions. |
| $k_1$ to $k_7$ | $\mathbb{R}^d$ | Coefficients of Parametric Transformation Functions |
| $T_{QOVR}$ | $\mathbb{R}^d$ | The transformation matrix scales the raw QOVR surface data to map the generated coordinate layers onto the baseline control points. |

The weight functions $w_i(u)$ introduced in (2) are defined using Bernstein polynomials (3)–(7) [10]. The general formulation of an $n^{th}$-degree Bernstein basis function under the parameter $u$ is shown as (3):

$$B_{i,n}(u) = \binom{n}{i} u^i (1-u)^{n-i} \quad i = 0, \dots, n. \quad (3)$$

To satisfy the partition of unity property for geometric stability, the sum of the basis functions equals one at any given parameter step, as stated in (4). Setting $n = 2$ defines the quadratic blending layers. The explicit definitions of the individual weight functions are structured in (5)–(7):

$$\sum_{i=0}^{n} B_{i,n}(u) = 1, \tag{4}$$

$$w_0(u) = B_{0,2}(u) = (1-u)^2, \tag{5}$$

$$w_1(u) = B_{1,2}(u) = 2u(1-u), \tag{6}$$

$$w_2(u) = B_{2,2}(u) = u^2. \tag{7}$$

Substituting these quadratic functions into the segment equation presents the blending architecture. The functional dependence of the independent parabolas is updated to reflect the parameter $u$, as shown in (8):

$$C(u) = w_0(u) * Q_0\big(r(u)\big) + w_1(u) * Q_1\big(s(u)\big) + w_2(u) * Q_2\big(t(u)\big). \tag{8}$$

To control the spatial adaptation without increasing the global polynomial degree of the spline, the internal parametric knots undergo direction-dependent transformations. The mapping operations that link the global variable $u$ to the localized domains defined in Table 1 are expressed in (9) and (10):

$$r_u = r(u) \leftarrow u, \tag{9}$$

$$s_u = s(u) \leftarrow u, \tag{10}$$

$$t_u = t(u) \leftarrow u. \tag{11}$$

By substituting these transformation functions into the active segment definition, the matrix-driven coupling framework is structured. However, to maintain alignment with the single-parameter notation before the spatial surface expansion, the global expression is shown in (12):

$$C(t) = w_0(u) * Q_0(r_u) + w_1(u) * Q_1(s_u) + w_2(u) * Q_2(t_u). \tag{12}$$

The spatial trajectory of the independent parabola $Q_0(r)$ is expressed in matrix form by factoring the local vector against the matrix coefficient block $B$, as stated in (13). To align the boundary constraints over the control points $P_0, P_1,$ and $P_2$ under the knot parameter $r_{ip}$ defined in Table 1, the structural network is mapped via (13):

$$Q_0(r) = \begin{bmatrix} r^2 & r & 1 \end{bmatrix} . B, \tag{13}$$

$$\begin{bmatrix} P_0 \\ P_1 \\ P_2 \end{bmatrix} = \begin{bmatrix} 0 & 0 & 1 \\ r_{ip}^2 & r_{ip} & 1 \\ 1 & 1 & 1 \end{bmatrix} . B, \tag{14}$$

By substituting the initial localized coefficients into a compact matrix notation $M_0$, the boundary system is structured through (15). To extract the algebraic basis matrix block $B$ directly from the control nodes, the linear system is inverted as shown in (16):

$$\begin{bmatrix} P_0 \\ P_1 \\ P_2 \end{bmatrix} = M_0 . B, \tag{15}$$

$$B = M_0^{-1}.\begin{bmatrix} P_0 \\ P_1 \\ P_2 \end{bmatrix}, \tag{16}$$

The analytical derivation of the inverse blending matrix $M_0^{-1}$ gives continuous algebraic transformation entries detailed in (17). Substituting this inverted kernel back into the initial polynomial description yields the spatial definition of the first independent parabola $Q_0(r)$ in (18):

$$M_0^{-1} = \begin{bmatrix} \frac{1}{r_{ip}} & \frac{1}{r_{ip}(r_{ip}-1)} & \frac{1}{1-r_{ip}} \\ \frac{-(r_{ip}+1)}{r_{ip}} & \frac{1}{r_{ip}(1-r_{ip})} & \frac{r_{ip}}{r_{ip}-1} \\ 1 & 0 & 0 \end{bmatrix}, \tag{17}$$

$$Q_0(r) = [r^2 \quad r \quad 1].M_0^{-1}.\begin{bmatrix} P_0 \\ P_1 \\ P_2 \end{bmatrix}. \tag{18}$$

Following the same matrix inversion sequence over the remaining control points $(P_1 \dots P_4)$ under their respective parameters $s$ and $t$, the formulations for the second and third independent parabolas, $Q_1(s)$ and $Q_2(t)$, are presented via (19) and (20):

$$Q_1(s) = [s^2 \quad s \quad 1].M_1^{-1}.\begin{bmatrix} P_1 \\ P_2 \\ P_3 \end{bmatrix}. \tag{19}$$

$$Q_2(t) = [t^2 \quad t \quad 1].M_2^{-1}.\begin{bmatrix} P_2 \\ P_3 \\ P_4 \end{bmatrix}. \tag{20}$$

The corresponding inverse blending matrices $M_1^{-1}$ and $M_2^{-1}$, which define the algebraic structure of these remaining segments based on the internal knot bounds $s_{ip}$ and $t_{ip}$, are expressed in (21)–(23):

$$M_1^{-1} = \begin{bmatrix} \frac{1}{s_{ip}} & \frac{1}{s_{ip}(s_{ip}-1)} & \frac{1}{1-s_{ip}} \\ \frac{-(s_{ip}+1)}{s_{ip}} & \frac{1}{s_{ip}(1-s_{ip})} & \frac{s_{ip}}{s_{ip}-1} \\ 1 & 0 & 0 \end{bmatrix}. \tag{21}$$

$$M_2^{-1} = \begin{bmatrix} \frac{1}{t_{ip}} & \frac{1}{t_{ip}(t_{ip}-1)} & \frac{1}{1-t_{ip}} \\ \frac{-(t_{ip}+1)}{t_{ip}} & \frac{1}{t_{ip}(1-t_{ip})} & \frac{t_{ip}}{t_{ip}-1} \\ 1 & 0 & 0 \end{bmatrix}. \tag{22}$$

By mapping the local parameter assignments $r_u, s_u,$ and $t_u$ directly into the consolidated curve framework, the expanded single-parameter QOVR equation is established in (23):

$$C(u) = (1-u)^2 [r_u^2 \quad r_u \quad 1] \cdot M_0^{-1} \cdot \begin{bmatrix} P_0 \\ P_1 \\ P_2 \end{bmatrix} + 2u(1-u)[s_u^2 \quad s_u \quad 1] \cdot M_1^{-1} \cdot \begin{bmatrix} P_1 \\ P_2 \\ P_3 \end{bmatrix}$$

$$+u^2 [t_u^2 \quad t_u \quad 1] \cdot M_2^{-1} \cdot \begin{bmatrix} P_2 \\ P_3 \\ P_4 \end{bmatrix}. \quad (23)$$

The evaluation of the blending architecture requires the assignment of localized boundary knots. By mapping the specific spatial thresholds $r_{ip}, s_{ip},$ and $t_{ip}$ alongside the primary coefficient matrix $M$, the distinct polynomial tracking paths are defined. Consequently, the consolidated framework structured as (13)–(23) serves as the structural expression for the independent parabolas. Within this formulation layer, the parameters $r, s,$ and $u$ serve as the local coordinates to calculate the points.

To evaluate the global equations over the combined QOVR spline framework, the weights of the independent parabolas are computed relative to the state of the parameter u. This requirement necessitates a parameter transformation sequence. Continuous mapping functions are used to link global variables with localized boundaries, avoiding global degree elevation [3]. The mathematical framework and the calculation of three direction-dependent parameter transformations are detailed in (24)–(26).

$$r_u = r(u) = \left(\frac{1-r_{ip}}{u_{ip}}\right).u + r_{ip}, \quad (24)$$

$$s_u = s(u) = \left(\frac{s_{ip}-u_{ip}}{u_{ip}^2-u_{ip}}\right).u^2 + \left(1 - \left(\frac{s_{ip}-u_{ip}}{u_{ip}^2-u_{ip}}\right)\right).\mathrm{u}, \quad (25)$$

$$t_u = t(u) = \left(\frac{t_{ip}}{1-u_{ip}}\right).u - \left(\frac{u_{ip}.t_{ip}}{1-u_{ip}}\right). \quad (26)$$

*3.2. Tensor Product Synthesis*

The spatial surface geometry is constructed by executing a tensor product between the orthogonal blending function matrices mapped along the parametric directions $u$ and $v$. This algebraic synthesis couples horizontal and vertical spline components over a static coordinate network. The mesh structure is systematically organized, as shown in Figure 2, to evaluate the spatial behavior of interconnected surface segments under local parameter variations.

The structural scheme in Figure 2 demonstrates that continuous mapping over the grid coordinates aligns the directional blending knots. Based on the orthogonal arrangement of the horizontal components $C_i(u)$ and vertical components $K_j(v)$, the spatial system constructs a unified coordinate mesh. To formalize the localized tracking expressions into an algebraic model suitable for computational scaling, the symmetric tensor product derivations are sequentially developed in (27)–(44).

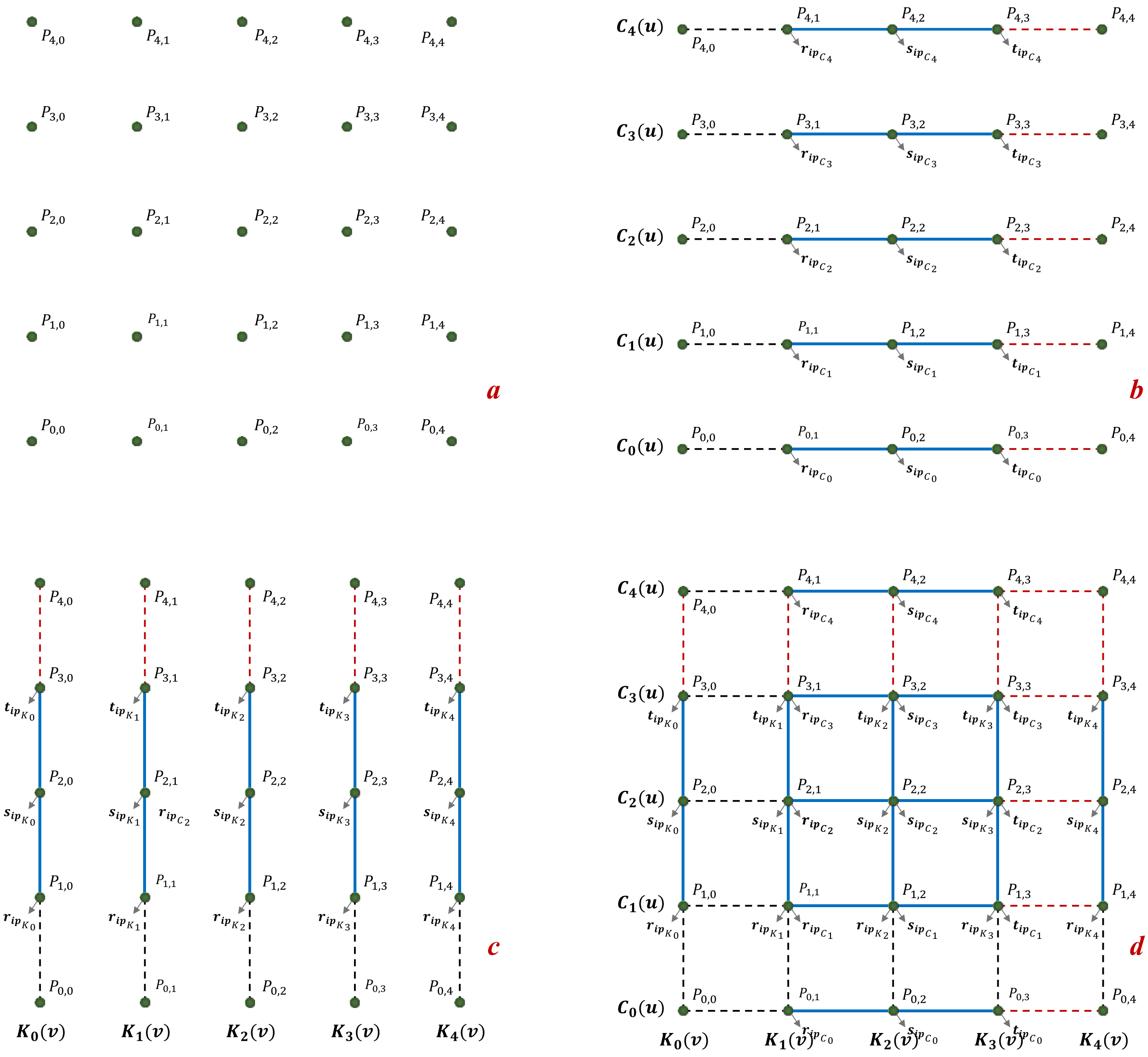

**Figure 2.** QOVR surface structural schema: **(a)** $5 \times 5$ control point matrix, **(b)** horizontal blending components $C_i(u)$, **(c)** vertical blending components $K_j(v)$, **(d)** composite surface grid under localized knot domains $C_i(u) \times K_j(v)$.

The overall geometric behavior of the QOVR surface is defined by combining the orthogonal parametric fields. The parametric tracking function that structures the surface coordinates through a double summation framework is given by (27):

$$QOVR(u, v) = \sum_{i=0}^{4} \sum_{j=0}^{4} P_{i,j} \cdot C_i(u) \cdot K_j(v) \tag{27}$$

To compute the vertical spline framework for the tensor product, the component $K(v)$ is formulated as a weighted linear combination of the directional independent parabolas $Q_j$, as shown in (28).

$$K(v) = w_0 . Q_0 + w_1 . Q_1 + w_2 . Q_{2,} \tag{28}$$

To format this vertical algebraic expression into a row-by-column matrix multiplication scheme, the linear layers are converted into a single-row vector product via (29):

$$K(v) = [w_0 \quad w_1 \quad w_2] \cdot \begin{bmatrix} Q_0 \\ Q_1 \\ Q_2 \end{bmatrix}, \tag{29}$$

By tracking the indices of the vector multiplication layers, the matrix representation of the vertical spline component is generalized using a summation symbol, as expressed in (30):

$$K(v) = \sum_{i=0}^{2} w_i \cdot Q_i. \tag{30}$$

To synchronize the orthogonal spline trajectories, the horizontal parameter assignments are mapped along the vertical axis relative to the parameter $v$, defined by the continuous parametric transformation functions shown in (31)–(33):

$$r_v = r(v) \leftarrow v, \tag{31}$$

$$s_v = s(v) \leftarrow v, \tag{32}$$

$$t_v = t(v) \leftarrow v. \tag{33}$$

The coordinates produced by these mapping functions are confined within localized parametric bounds to maintain cross-directional continuity. To define these spatial domains under the static network knots, the separate internal knot vectors for the horizontal and vertical spline layers are structured via (34) and (35):

$$r_{ip_{C_i}}, s_{ip_{C_i}}, u_{ip_i}, t_{ip_{C_i}}, \tag{34}$$

$$r_{ip_{K_j}}, s_{ip_{K_j}}, v_{ip_j}, t_{ip_{K_j}}. \tag{35}$$

To formulate the blending weights relative to the parameter steps, the spatial grid invokes coordinate transformation functions. The mapping segments that transform the parameter $u$ into the horizontal knot intervals are calculated as shown in (36)–(38):

$$r_u = r(u) = \left(\frac{1-r_{ip_C}}{u_{ip}}\right).u + r_{ip_C}, \tag{36}$$

$$s_u = s(u) = \left(\frac{s_{ip_C}-u_{ip}}{u_{ip}^2-u_{ip}}\right).u^2 + \left(1-\left(\frac{s_{ip_C}-u_{ip}}{u_{ip}^2-u_{ip}}\right)\right).t, \tag{37}$$

$$t_u = t(u) = \left(\frac{t_{ip_C}}{1-u_{ip}}\right).u - \left(\frac{u_{ip}.t_{ip_C}}{1-u_{ip}}\right). \tag{38}$$

Symmetrically, the functions deployed to transform the parameter $v$ into the localized vertical knot segments are shown in (39)–(41):

$$r_v = r(v) = \left(\frac{1-r_{ip_K}}{v_{ip}}\right).v + r_{ip_K}, \tag{39}$$

$$s_v = s(v) = \left(\frac{s_{ip_K}-v_{ip}}{v_{ip}^2-v_{ip}}\right).v^2 + \left(1-\left(\frac{s_{ip_K}-v_{ip}}{v_{ip}^2-v_{ip}}\right)\right).v, \tag{40}$$

$$t_v = t(v) = \left(\frac{t_{ip_K}}{1-v_{ip}}\right).n - \left(\frac{v_{ip}.t_{ip_K}}{1-v_{ip}}\right). \tag{41}$$

To structure the algebraic modeling of the surface framework, the spline formulations are extended into matrix-driven vector expressions. The expanded tracking sequence for the horizontal spline component $C_i(u)$ mapping over the specific row points of the control mesh is stated in (42):

$$\begin{gathered} C_i(u) = w_0(u) \cdot \begin{bmatrix} r_{u_i}^2 & r_{u_i} & 1 \end{bmatrix} \cdot M_{0C_i}^{-1} + w_1(u) \cdot \begin{bmatrix} s_{u_i}^2 & s_{u_i} & 1 \end{bmatrix} \cdot M_{1C_i}^{-1} \\ +w_2(u) \cdot \begin{bmatrix} t_{u_i}^2 & t_{u_i} & 1 \end{bmatrix} \cdot M_{2C_i}^{-1}, \end{gathered} \tag{42}$$

Symmetrically, the inverted blending layers and transformation functions are applied along the vertical parameter direction. The complete vector formulation for the directional spline component $K_j(v)$ mapping over the column points is expressed in (43):

$$\begin{gathered} K_j(v) = w_0(v) \cdot \begin{bmatrix} r_{v_j}^2 & r_{v_j} & 1 \end{bmatrix} \cdot M_{0K_j}^{-1} + w_1(v) \cdot \begin{bmatrix} s_{v_j}^2 & s_{v_j} & 1 \end{bmatrix} \cdot M_{1K_j}^{-1} \\ +w_2(v) \cdot \begin{bmatrix} t_{v_j}^2 & t_{v_j} & 1 \end{bmatrix} \cdot M_{2K_j}^{-1}. \end{gathered} \tag{43}$$

By executing the tensor product across all active blending segments, the double summation defined by (27) is expanded. The consolidated expanded algebraic equation representing the global coordinate fields of the QOVR surface is shown in (44):

$$\begin{gathered} QOVR(u,v) = \sum_{i=0}^{4} \sum_{j=0}^{4} P_{i,j}\, C_i(u) K_j(v) = \\ P_{0,0}C_0(u)K_0(v) + P_{0,1}C_0(u)K_1(v) + \cdots + P_{0,4}C_0(u)K_4(v) \\ +P_{1,0}C_1(u)K_0(v) + P_{1,1}C_1(u)K_1(v) + \cdots + P_{1,4}C_1(u)K_4(v) \\ +P_{2,0}C_2(u)K_0(v) + P_{2,1}C_2(u)K_1(v) + \cdots + P_{2,4}C_2(u)K_4(v) \\ +P_{3,0}C_3(u)K_0(v) + PQ_{3,1}C_3(u)K_1(v) + \cdots + P_{3,4}C_3(u)K_4(v) \\ +P_{4,0}C_4(u)K_0(v) + P_{4,1}C_4(u)K_1(v) + \cdots + P_{4,4}C_4(u)K_4(v). \end{gathered} \tag{44}$$

The formulation steps developed through (27)–(44) close this subsection by defining the algebraic structure of the QOVR surface. By expanding the tensor product into matrix operations, the coordinate field is mapped across the static grid segments. This mathematical framework provides the equations used for the subsequent differential analyses and smoothness evaluations at the interface zones.

### *3.3. Analytical Continuity Evaluations Across Local Boundaries*

To verify the spatial integrity of the QOVR surface framework within a CAGD environment, the boundary interfaces must satisfy local closure conditions. The analytical evaluations check the continuity states across both the exterior perimeters and the shared interior junctions governed by direction-dependent parameter transformations.

**Lemma 1.** *($C^0$ Continuity-Position Alignment).*

Zero-order position closure requires that the coordinates along the shared interface between adjacent patches yield identical geometric tracks. For two neighboring patches meeting at a shared internal network line where the local parameter $u$ or $v$ reaches its structural boundary limit, the analytical boundary condition is verified by computing the left-hand and right-hand limit expressions:

$$\lim_{u \to 1^-} QOVR_{left}(u, v) = \lim_{u \to 0^+} QOVR_{right}(u, v). \tag{45}$$

To verify this property, the mathematical limits are evaluated at the shared interior network junction where $u \to 1^-$ and $u \to 0^+$. The left-hand surface equation $QOVR_{left}(u, v)$ approaching the junction from the domain $u \in [0.0, 1.0)$ utilizes the terminal knot boundaries. Substituting $u = 1.0$ into the horizontal spline matrix kernels $C_i(u)$ defines the boundary state:

$$C_i(1^-) = [0 \quad 0 \quad 1 \quad 0 \quad 0]_i, \tag{46}$$

This mapping projects the spatial coordinates directly onto the central control points column:

$$\lim_{u \to 1^-} Q\,OVR_{left}(u, v) = \sum_{j=0}^{4} P_{2,j} \cdot K_j(v), \tag{47}$$

Symmetrically, the right-hand surface equation $QOVR_{right}(u, v)$ entering the adjacent patch domain where $u \in (0.0, 1.0]$ relies on the initial parameter threshold $u = 0.0$, yielding the horizontal kernel array state (46), (48):

$$C_i(0^+) = [0 \quad 0 \quad 1 \quad 0 \quad 0]_i, \tag{48}$$

Evaluating this initial boundary identifies the identical control node line across the shared intersection:

$$\lim_{u \to 0^+} Q\,OVR_{right}(u, v) = \sum_{j=0}^{4} P_{2,j} \cdot K_j(v). \tag{49}$$

Comparing (47) and (49) establishes that both spatial limits converge to the identical coordinate expression, verifying (45) under static coordinate networks.

**Lemma 2.** *($C^1$ Continuity-Tangent Alignment).*

First-order continuity requires that the partial derivative vectors across the shared interior boundaries maintain directional alignment without sudden coordinate steps to support smoothness inside interactive geometric environments. The partial derivative tracking along the parameter axes is expressed in (50):

$$\lim_{u \to 1^-} \frac{\partial QOVR_{left}(u,v)}{\partial u} = \lim_{u \to 0^+} \frac{\partial QOVR_{right}(u,v)}{\partial u}. \tag{50}$$

Differentiating the horizontal kernels with respect to $u$ and computing the limit step at the terminal boundary $u = 1.0.$ yields the left-hand derivative weights shown in (51):

$$\frac{dC_i(1^-)}{du} = [0 \quad -0.5 \quad 0 \quad 0.5 \quad 0]_i, \tag{51}$$

Substituting these differential weights into the spatial tensor operations structures the left-hand derivative vector expression:

$$\lim_{u \to 1^-} \frac{\partial QOVR_{left}(u,v)}{\partial u} = \sum_{j=0}^{4} \left(-0.5 \cdot P_{1,j} + 0.5 \cdot P_{3,j}\right) \cdot K_j(v), \tag{52}$$

For the right-hand surface equation, differentiating the initial blending functions and substituting the starting parameter threshold $u = 0.0$ maps the initial derivative kernel layout:

$$\frac{dC_i(0^+)}{du} = [0 \quad -0.5 \quad 0 \quad 0.5 \quad 0]_i, \tag{53}$$

Applying these initial weights onto the shared coordinate schema gives the right-hand partial derivative expression:

$$\lim_{u \to 0^+} \frac{\partial QOVR_{right}(u,v)}{\partial u} = \sum_{j=0}^{4} \left(-0.5 \cdot P_{1,j} + 0.5 \cdot P_{3,j}\right) \cdot K_j(v). \tag{54}$$

The algebraic identity shown in (52)–(54) satisfies the first-order tangent vector continuity limits, confirming that the tangent plane remains smooth across the shared junctions without adding global polynomial degree elevation.

To evaluate higher-order boundary behavior, the second-order partial derivative formulations are computed directly from the blending matrices. Differentiating the horizontal kernels twice and substituting the terminal limit $u = 1.0$ yields the left-hand quadratic differential weights (55):

$$\frac{d^2C_i(1^-)}{du^2} = [2 \quad -4 \quad 2 \quad 0 \quad 0]_i, \tag{55}$$

This maps the left-hand second-order partial derivative expression shown in (56):

$$\lim_{u \to 1^-} \frac{\partial^2 QOVR_{left}(u,v)}{\partial u^2} = \sum_{j=0}^{4} \left(2 \cdot P_{0,j} - 4 \cdot P_{1,j} + 2 \cdot P_{2,j}\right) \cdot K_j(v), \tag{56}$$

Evaluating the initial second-order derivative layout at the starting parameter threshold $u = 0.0$ gives (57):

$$\frac{d^2C_i(0^+)}{du^2} = [0 \quad 0 \quad 2 \quad -4 \quad 2]_i, \tag{57}$$

The resulting right-hand second-order partial derivative expression is stated in (58):

$$\lim_{u \to 0^+} \frac{\partial^2 QOVR_{right}(u,v)}{\partial u^2} = \sum_{j=0}^{4} \left(2 \cdot P_{2,j} - 4 \cdot P_{3,j} + 2 \cdot P_{4,j}\right) \cdot K_j(v), \tag{58}$$

The algebraic evaluation of the cross-boundary second derivatives indicates that $C^2$ curvature continuity is not maintained across the shared node interfaces. Differentiating the matrix framework twice yields the respective quadratic differential weights at the boundaries. Due to the inherent quadratic properties of the baseline blending kernels and the independent control point configurations of the adjacent segments, the left-hand and right-hand second-order limits yield unequal values. This mathematical boundary mismatch represents a structural property derived directly from the localized knot constraints. The boundary conditions and algebraic limits along the $v$ direction are determined by the blending kernels $K_j(v)$. Differentiating the tensor-product formulation with respect to the $v$ axis defines the cross-boundary expressions at $v = 0$ and $v = 1$. Evaluating these positional limits across the node interfaces yields identical coordinate vectors, which verifies $C^0$ continuity. For the first-order derivatives, the evaluation yields identical tangent weights, satisfying $C^1$ continuity along the $v$ axis. Finally, evaluating the second-order derivatives produces unequal matrix weights at the nodes, confirming that $C^2$ continuity is not maintained along the transverse direction.

*3.4. Homogeneous Transformation Matrix and Coordinate Alignment*

To position the structured QOVR surface within the global spatial environment, the localized network coordinates undergo affine transformations. The geometric framework executes these spatial modifications by deploying a $4 \times 4$ homogeneous transformation matrix. The unified algebraic representation of the homogeneous transformation operator is given in (59)–(61):

$$T = \begin{bmatrix} r_{11} & r_{12} & r_{13} & t_x \\ r_{21} & r_{22} & r_{23} & t_y \\ r_{31} & r_{32} & r_{33} & t_z \\ 0 & 0 & 0 & 1 \end{bmatrix}. \tag{59}$$

$$QOVR_{Transformed} = T_{QOVR} * QOVR_{Raw}, \tag{60}$$

$$\begin{bmatrix} X_{transformed} \\ Y_{transformed} \\ Z_{transformed} \end{bmatrix} = T_{QOVR} \cdot \begin{bmatrix} X_{raw} \\ Y_{raw} \\ Z_{raw} \end{bmatrix}. \tag{61}$$

The evaluation of the unified geometric transformations requires a consistent notation for all local variables and coordinate transformations. The numerical limits, coordinate definitions, and structural parameter domains, structured within a unified matrix, are summarized in Table 1 for the spline and extended surface frameworks.

## 4. Algorithmic Framework and Software Implementation

This section presents computer graphics algorithms and software implementation framework utilized to generate the spatial network. The data routing pipeline aligns the coordinate transformation functions, spline matrix kernels, and boundary conditions through a processing pipeline. To transition from the algebraic surface formulations into a visual verification tool, the execution sequence is documented using flowchart schematics, algorithmic steps, and graphical interface components. This systematic tracking checks the algorithm execution logic under static control point arrays before deploying the final data paths.

### 4.1. Software Implementation and Execution Logic

The operational flow of the spatial framework is organized as a sequential processing pipeline to map the control coordinates onto the surface layer. This pipeline handles the calculations by moving from initial data inputs through parameter transformations, matrix multiplications, and boundary continuity routines. To visualize these computational steps and the execution order within the software routine, the underlying logic is mapped in Figure 3.

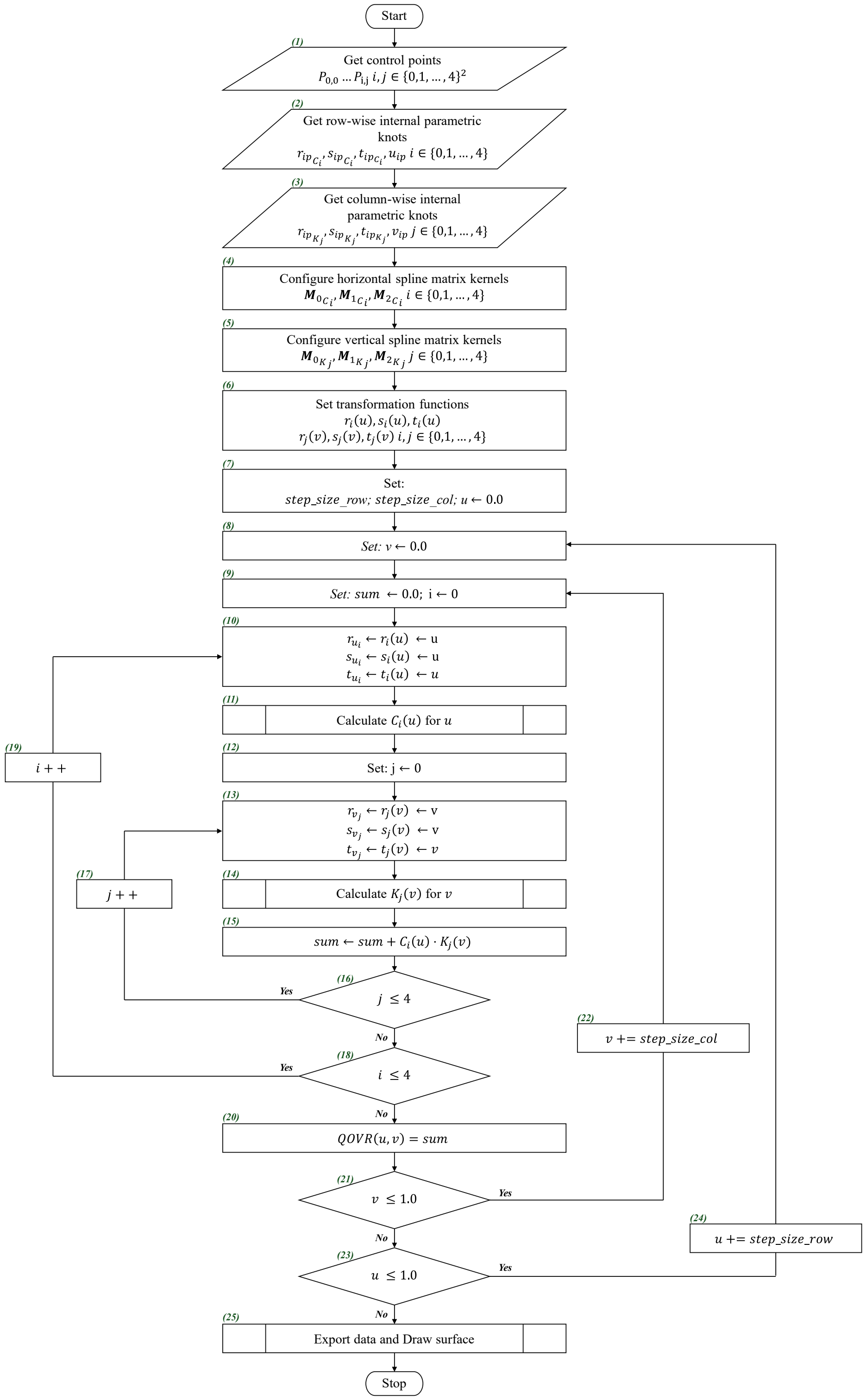

**Figure 3.** Algorithmic flowchart of the QOVR surface framework.

Figure 3 illustrates the surface generation process, starting with reading control points and internal knots per Table 1. Steps (4) and (5) configure directional inverse blending matrices, while Step (6) initializes transformation functions for spatial mapping at local boundaries.

The computational process uses nested loops to evaluate the surface coordinates, where steps (7)–(9) initialize the parametric segments and steps (10)–(13) apply the transformations across the variables $u$ and $v$. The directional splines are calculated in steps (11) and (14), combined via a summation framework in step (15), and verified in steps (16) and (18). This pipeline leads to the surface resolution in step (20), boundary mapping in steps (21) and (23), and the homogeneous transformation in step (25) prior to coordinating export in step (26).

To translate the sequential blocks illustrated in Figure 3 into a direct code implementation, the execution steps are structured as a formal instruction template. This logic translates the nested network evaluations, matrix multiplications, and domain step controls into computational steps, with the pseudocode implementation template that defines the program loops and parameter mapping sequence presented in Algorithm 1.

**Algorithm 1:** Computational Pipeline for QOVR Surface Generation.

---

Input: Control points vector $P_{\{i,j\}},\ i,j \in \{0,1,\dots,4\}^2$

Row-wise internal parametric knots $r_{ipC_i}, s_{ipC_i}, u_{ip_i}, t_{ipC_i} \in [0.0, 1.0];\ i \in \{0,1,\dots,4\}$

Column-wise internal parametric knots $r_{ipK_j}, s_{ipK_j}, v_{ip_j}, t_{ipK_j} \in [0.0, 1.0];\ j \in \{0,1,\dots,4\}$

Step sizes (step_size_row, step_size_col)

Output: QOVR surface raw and transformed coordinates, Bounding Boxes, and Transformation Matrix ($T_{QOVR}$) structured in data array

---

1: Configure horizontal spline matrix kernels ($M_{0C_i}, M_{1C_i}, M_{2C_i}\ i \in \{0,1,\dots,4\}$)
2: Configure vertical spline matrix kernels ($M_{0K_j}, M_{1K_j}, M_{2K_j}\ j \in \{0,1,\dots,4\}$)
3: Set transformation functions $r_i(u), s_i(u), t_i(u),\ r_j(v), s_j(v), t_j(v)\ \ i,j \in \{0,1,\dots,4\}$
4: Set $u \leftarrow 0.0$
5: While $u \le 1.0$ do
6: Set $v \leftarrow 0.0$
7: While $v \le 1.0$ do
8: Set $sum \leftarrow 0.0, i \leftarrow 0$
9: While $i \le 4$ do
10: Set dynamically transformed horizontal parameters,
11: $r_{u_i} \leftarrow r_i(u);\ s_{u_i} \leftarrow s_i(u);\ t_{u_i} \leftarrow t_i(u)$ using (36)–(38)
12: Calculate $C_i(u)$ for $u$ (42)
13: Set $j \leftarrow 0$
14: While $j \le 4$ do
15: Set dynamically transformed vertical parameters,
16: $r_{v_j} \leftarrow r_j(v);\ s_{v_j} \leftarrow s_i(v);\ t_{v_j} \leftarrow t_i(v)$ using (39)–(41)
17: Calculate $K_j(v)$ for $v$ (43)
18: $sum \leftarrow sum + C_i(u) \cdot K_j(v)$
19: $j \leftarrow j + 1$
20: End While

---

```
        i ← i + 1
        End While
        QOVR(u, v) ← sum  (44)
        v ← v + step_size_col
    End While
    u ← u + step_size_row
End While
Export surface coordinate data and Draw surface
```

The workflow from Figure 3 to Algorithm 1 defines the execution steps of the QOVR model. Structuring the logic through loop boundaries and knot thresholds maps the data paths across the mesh intervals. To observe the behavior of these instruction blocks under parameter modifications, the source code is integrated into a user interface, presented in the next subsection.

### *4.2. Functional Modules of the Visual Interface*

The software routine is executed within a graphical interface to verify the numerical outputs of the surface equations. This testing interface is divided into parametric input fields, matrix configuration inputs, and coordinate display panels. Users modify the knot boundaries and control mesh inputs directly through text fields and slider controls. The interface processes these input updates and reconstructs the spatial mesh coordinates across the defined boundaries. The design arrangement and operational control zones of the implemented environment are illustrated in Figure 4.

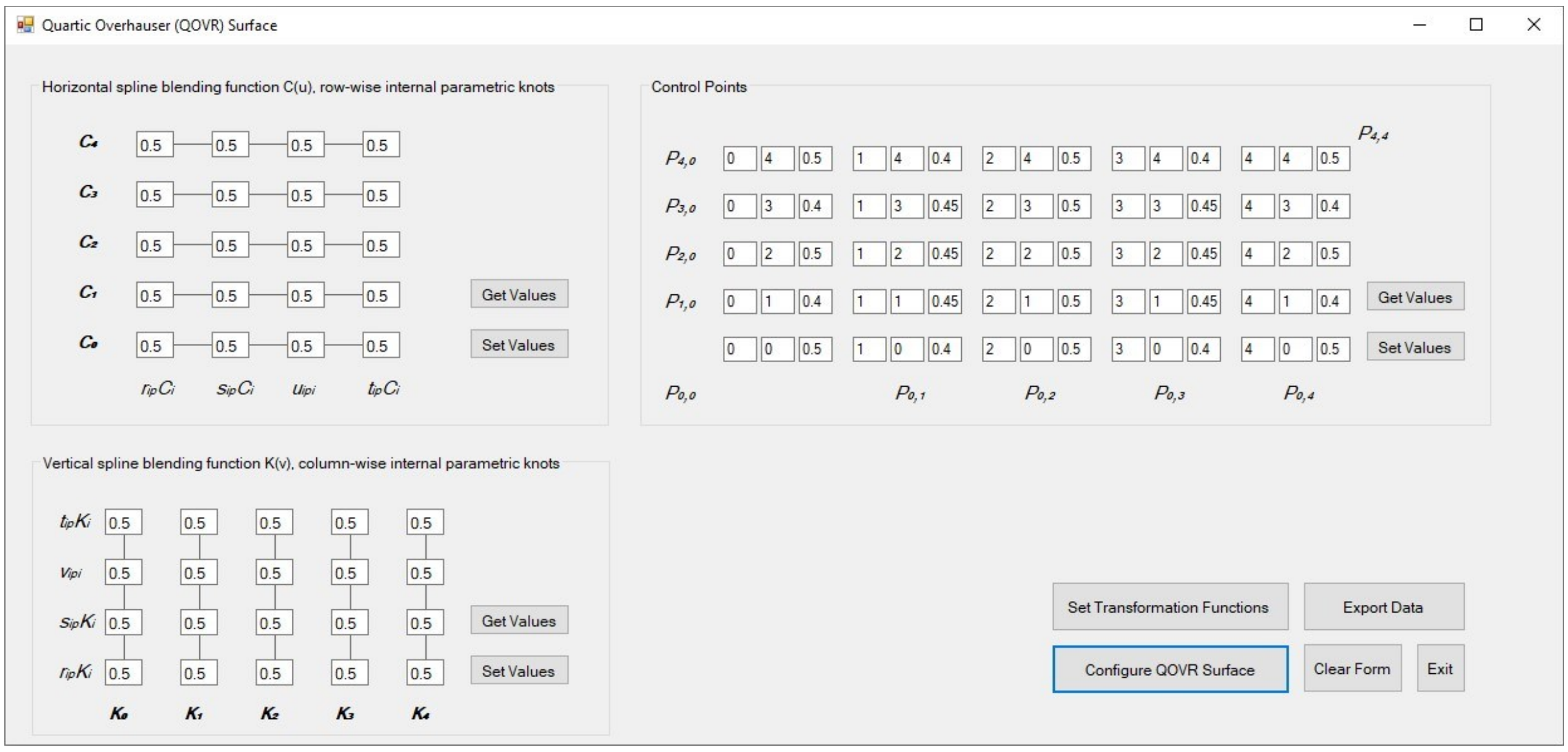

**Figure 4.** Layout arrangement of the GUI user controls and functional modules, featuring internal parametric knots and control points.

The interface layout illustrated in Figure 4 organizes the control parameters into distinct window blocks. The upper left quadrant contains the row-wise internal parametric knot configuration fields for the horizontal spline blending functions $C_i(u)$, presenting the thresholds $r_{ipC_i}, s_{ipC_i}, u_{ip_i}$, and $t_{ipC_i}$ across the curves $C_0$ to $C_4$. Symmetrically, the lower left quadrant contains the column-wise internal parametric knot fields for the vertical spline blending functions $K_j(v)$, presenting the variables $r_{ipK_j}, s_{ipK_j}, u_{ip_j},$ and $t_{ipK_j}$ for the vectors $K_0$ to $K_4$. Both directional knot layers integrate "Get Values" and "Set Values" command buttons to read and write active boundary states.

The right-hand side of the control interface is divided between matrix data entry zones and operational execution triggers. The upper right window manages the spatial control network coordinates through a $5 \times 5$ input matrix, structuring the points from the base element $P_{0,0}$ up to the terminal point $P_{4,4}$. This control matrix is equipped with independent "Get Values" and "Set Values" actions.

The lower right control panel contains the execution modules for the computational pipeline. Triggering the "Configure QOVR Surface" command executes Step (25) of the algorithmic steps within the background software routine. This execution sequence enforces a data dependency model where the raw surface coordinate data must be fully generated before the transformation matrix can be configured. Consequently, the computational pipeline executes the initial surface coordinates to build the raw surface data layer. The software then computes the operational values of the transformation matrix to execute the transformed surface coordinates. To provide data paths for differential checking, bounding box validation, and external visualization, the routine exports the control points, internal parametric knots, bounding boxes, transformation matrix elements, raw surface data, and transformed surface data during the export routine controlled by the "Export Data" command.

## 5. Geometric Implementation and Boundary Analysis

This section presents graphical results, numerical configurations, and software execution outputs designed to evaluate the spatial behavior of the QOVR surface framework within dynamic design spaces. The computational pipeline reads discrete input point matrices to execute the algebraic transformation functions and matrix operations mapped along orthogonal coordinate axes. To analyze the model responses, the implementation pipeline evaluates the coordinate layers across distinct geometric test cases under varying knot transformations and boundary conditions. To examine the analytical continuity and localized shape modifications presented in the theoretical model, the generated data points are evaluated sequentially through surface network visualization, bounding volume enclosures, and shape control sensitivity metrics. This systematic evaluation monitors the boundary states and derivative limits across the surface network interfaces.

### *5.1. Surface Mesh Generation and Transformation Operations*

To analyze the physical manifestation of the algebraic surface equations within the graphical interface routine, the operational software processes the discrete control points arrays under varied transformation steps. The computational pipeline reads the baseline matrix data to reconstruct the spatial coordinates before deploying the homogeneous transformation matrix. This numerical framework applies directional scaling adjustments alongside spatial translation factors, providing the basis to examine whether the coordinate alignment operations maintain the underlying spline continuity status at the patch interfaces. To examine the operational sequencing order by comparing the untransformed coordinates with the post-generation affine outputs, the spatial networks are evaluated as illustrated in Figure 5.

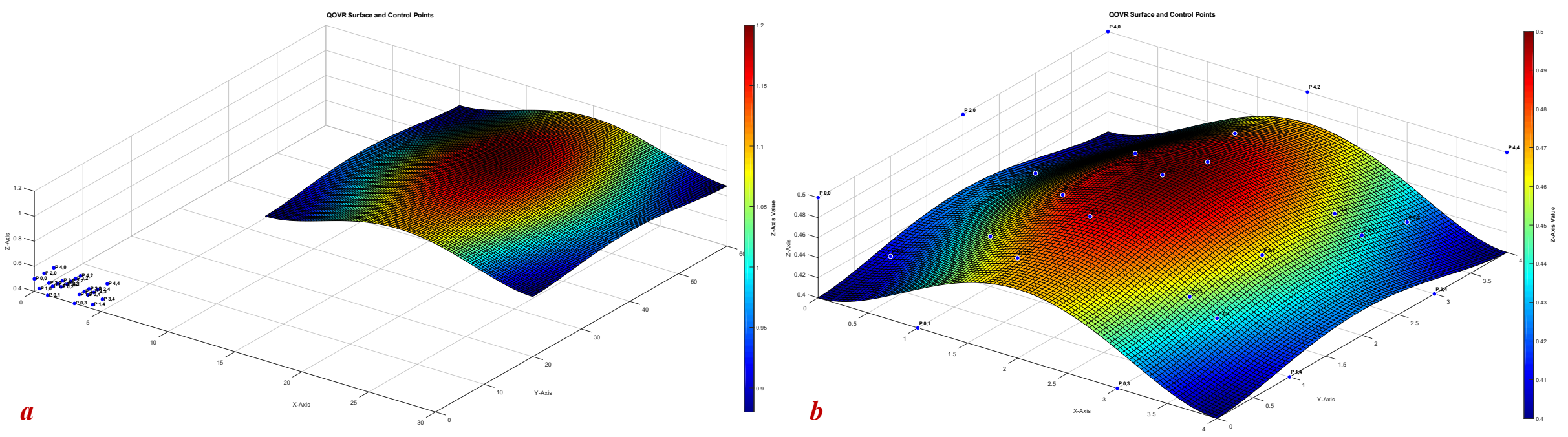

**Figure 5.** Operational arrangement of the generated geometric entities: (**a**) raw surface data and spatial control points matrix configuration, (**b**) transformed QOVR surface mesh after applying the homogeneous matrix operators.

To align the generated coordinate layers with the target display constraints, the transformation parameters along the orthogonal axes are computed through linear equations based on the extracted spatial bounds. The computational pipeline maps the surface bounding box (BB) intervals directly onto the target control points limits. When the internal parametric knots are set to 0.5 along the orthogonal directions, the bounding box (BB) values for the control points and the raw QOVR surface coordinates are defined as follows:

$$BB_{CntlP} = \begin{bmatrix} 0 & 4 \\ 0 & 4 \\ 0.4 & 0.5 \end{bmatrix}, BB_{QOVR_{raw}} = \begin{bmatrix} 10 & 30 \\ 20 & 60 \\ 0.88 & 1.2 \end{bmatrix}.$$

For the x-axis calculation, the transformation maps the initial boundary within the interval $X_{raw} = 10 \to X_{transformed} = 0 \; and \; X_{raw} = 30 \to X_{transformed} = 4$. The scaling slope $m_x$ and translation factor $n_x$ are evaluated as follows:

$$m_x = \frac{(4 - 0)}{(30 - 10)} = \frac{4}{20} = 0.2,$$
$$0 = 0.2 \cdot 10 + n_x \Rightarrow n_x = -2,$$

The linear equation for the horizontal mapping yields:

$$X_{Transformed} = 0.2 \cdot X_{Raw} - 2.$$

For the y-axis calculation, the transformation maps the initial boundary within the interval $Y_{raw} = 20 \to Y_{transformed} = 0 \; and \; Y_{raw} = 60 \to Y_{transformed} = 4$. The scaling slope $m_y$ and translation factor $n_y$ are evaluated as follows:

$$m_y = \frac{(4 - 0)}{(60 - 20)} = \frac{4}{40} = 0.1,$$
$$0 = 0.1 \cdot 20 + n_y \Rightarrow n_y = -2,$$

The linear equation for the lateral mapping gives:

$$Y_{transformed} = 0.1 \cdot Y_{raw} - 2.$$

For the z-axis calculation, the transformation maps the initial boundary within the interval $Z_{raw} = 0.88 \rightarrow Z_{transformed} = 0.4 \, and \, Z_{raw} = 1.20 \rightarrow Z_{transformed} = 0.5$. The scaling slope $m_z$ and translation factor $n_z$ are evaluated as follows:

$$m_z = \frac{(0.5 - 0.4)}{(1.20 - 0.88)} = \frac{0.1}{0.32} = 0.3125,$$

$$0.4 = 0.3125 \cdot 0.88 + n_z \Rightarrow n_z = 0.4 - 0.275 = 0.125,$$

The linear equation for the vertical coordinate transformation yields the parametric relationship:

$$Z_{transformed} = 0.3125 \cdot Z_{raw} + 0.125.$$

The spatial bounds calculated across these orthogonal intervals, along with the untransformed coordinates, are provided for numerical analysis as specified in the Data Availability Statement. To implement the surface generation algorithm within a computer graphics framework, the discrete surface point cloud coordinates $(X_{raw}, Y_{raw}, Z_{raw})$ are mapped onto the control point mesh coordinate system $(X_{transformed}, Y_{transformed}, Z_{transformed})$. Instead of computing linear interpolation equations for each axis independently, the computational pipeline utilizes a $4 \times 4$ Homogeneous Transformation $Matrix\ (T_{QOVR})$. This representation executes scaling and translation operations through a single matrix-vector multiplication within the software routine. The transformation matrix incorporating the calculated normalization boundaries is expressed in (62).

$$T_{QOVR} = \begin{bmatrix} 0.2 & 0 & 0 & -2 \\ 0 & 0.1 & 0 & -2 \\ 0 & 0 & 0.3125 & 0.125 \\ 0 & 0 & 0 & 1 \end{bmatrix}. \quad (62)$$

The mapping operation for a surface point vector is executed as follows in (63):

$$\begin{bmatrix} X_{transformed} \\ Y_{transformed} \\ Z_{transformed} \\ 1 \end{bmatrix} = \begin{bmatrix} 0.2 & 0 & 0 & -2 \\ 0 & 0.1 & 0 & -2 \\ 0 & 0 & 0.3125 & 0.125 \\ 0 & 0 & 0 & 1 \end{bmatrix} \cdot \begin{bmatrix} X_{raw} \\ Y_{raw} \\ Z_{raw} \\ 1 \end{bmatrix}. \quad (63)$$

In the subsequent operational stages, alterations in the control point positions or updates to the internal parametric knot vectors trigger a recomputation sequence. The computational pipeline executes the raw QOVR surface geometry first, followed by the formulation of the corresponding homogeneous transformation matrix, verifying that all post-generation outputs are visualized at an identical spatial scale with the control network.

### *5.2. Bounding Volumes*

The limits of the generated coordinate layers are examined by evaluating the surface behavior under alternative internal parameter sets. This analysis processes a static control matrix while shifting the internal parametric knot vectors to analyze the geometric sensitivity of the network structure. Modifying the boundary thresholds determines how the coordinate layers expand or compress across the local zones, providing the mathematical data needed to monitor surface structural variations. To visualize the coordinate modifications caused by changing the internal knot vectors, the spatial networks are generated as illustrated in Figure 6.

The structural variations observed across the individual panels of Figure 6 demonstrate the geometric influence of the internal parameter states under a static control points matrix. Symmetrically setting all parametric thresholds to 0.5 in Figure 6a results in a uniform distribution across network layers and a balanced baseline curvature. Shifting the variables to the values in Figure 6b, $r_{ip} = 0.3, s_{ip} = 0.5, u_{ip} = 0.5$, and $t_{ip} = 0.7$, leads to an asymmetrical compression of the tracking layers toward the initial perimeter segments.

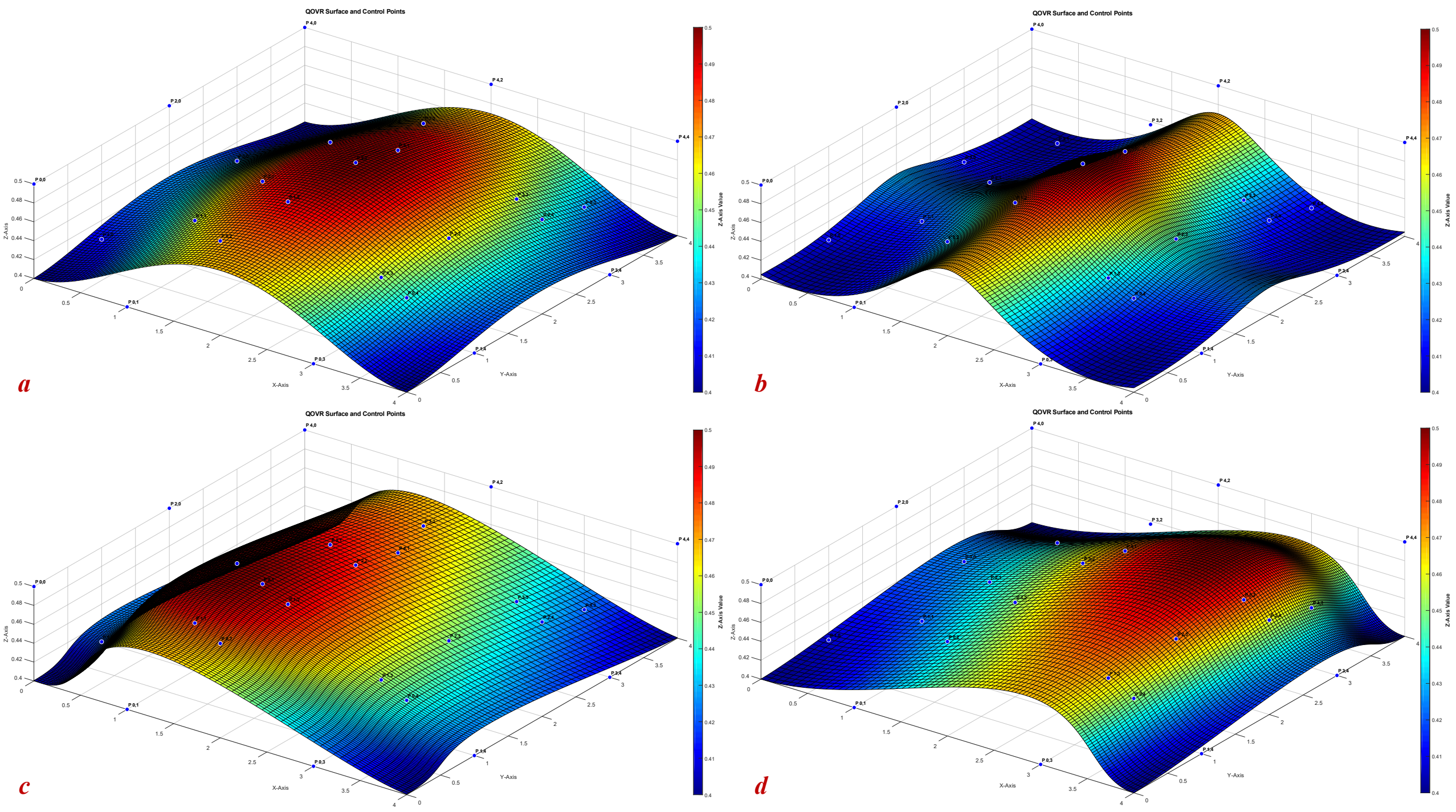

**Figure 6.** Surface variations under variable internal parametric knots for identical control points matrix scheme: **(a)** all parameter thresholds set to 0.5, **(b)** $\boldsymbol{r_{ip} = 0.3,\ s_{ip} = 0.5, u_{ip} = v_{ip} = 0.5,\ t_{ip} = 0.7}$, **(c)** $\boldsymbol{r_{ip} = 0.4,\ s_{ip} = 0.6, u_{ip} = v_{ip} = 0.7,\ t_{ip} = 0.4}$, **(d)** $\boldsymbol{r_{ip} = 0.5,\ s_{ip} = 0.4, u_{ip} = v_{ip} = 0.3,\ t_{ip} = 0.5}$.

Alternatively, the configuration shown in Figure 6c exhibits the opposite structural behavior by mapping the parameters to $0.4, 0.6, 0.7$, and $0.4$, respectively. This parameter shifts the peak coordinate scaling toward the terminal execution zones, thereby modifying the scheme alignment. Finally, the parameter configuration in Figure 6d, with bounds set at $0.5, 0.4, 0.3$, and $0.5$, compresses the internal tracking coordinates around the central network zones. The geometric variations observed between Figure 6a–d demonstrate that the localized tracking layers respond directly to internal parametric knot updates without requiring any global coordinate adjustments.

### *5.3. Shape Control Symmetry Sensitivity*

The response of the surface networks is evaluated by analyzing the coordinate distributions across alternative knot boundaries. This visualization evaluates spatial variations through color-mapped intensity fields to examine the geometric alignment and structural symmetry of the network intervals under variable internal parameters. Computing these spatial distributions reveals how the vertex updates expand or compress across the local zones, thereby executing the symmetry sensitivity analysis of the surface. To visualize the symmetric distribution fields corresponding to the configurations tested in Figure 6, the computational outputs are mapped as illustrated in Figure 7.

The intensity layers observed across the individual quadrants of Figure 7 illustrate the localized geometric variations under variable-parameter bounds. In Figure 7a, the symmetrical color distribution indicates a balanced network alignment, reflecting the geometric symmetry of the coordinate layers under uniform knot thresholds. When evaluating the variable distribution fields in Figure 7b, the intensity compresses along the central lines while expanding towards the outer boundary sectors, demonstrating a geometric transition that preserves the surface fairness without triggering wave-like artefacts.

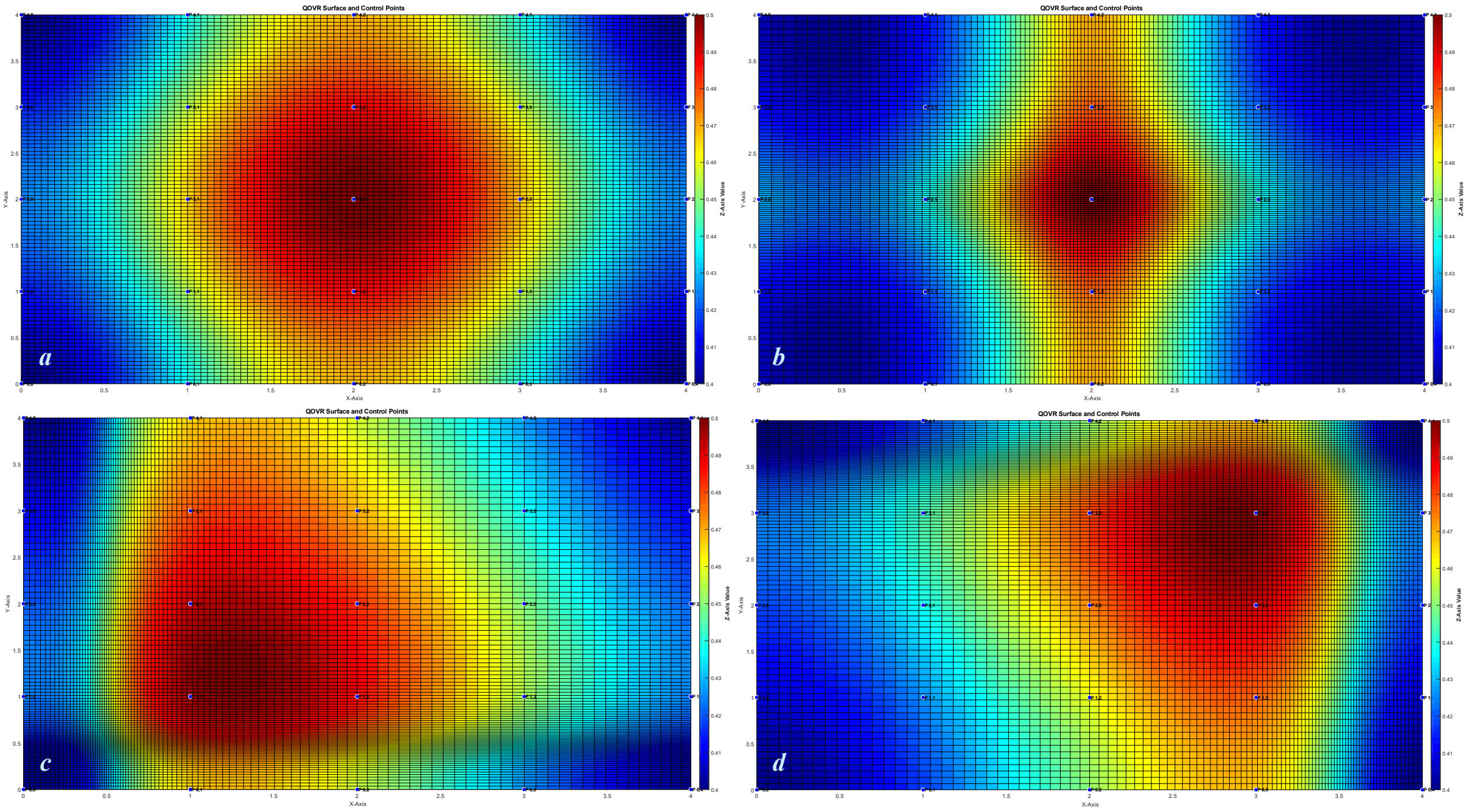

**Figure 7.** Coordinate intensity distribution fields and surface scheme variations evaluated for the identical parameter configurations tested from Figure 6a–d: **(a)** color-mapped intensity under uniform threshold limit set at 0.5, **(b)** distribution fields under asymmetric parameter tracking limits, **(c)** shifted coordinate compression layers focused toward the initial boundaries, **(d)** regional layout scaling updates tracking toward the terminal boundary limit.

Furthermore, shifting the thresholds to the configuration evaluated in Figure 7c yields a directional intensity consolidation focused on the initial boundary lines. Conversely, the parameter configuration structured in Figure 7d shifts the coordinate intensity towards the terminal boundary limits. Despite these regional shape modifications observed from Figure 7a – d, the outermost boundary coordinates across all configurations maintain positional invariance, showing that the transformed coordinate layers remain anchored to the baseline reference lines. This geometric alignment indicates that the localized coordinate layers satisfy positional closure under variable-parameter states, preventing edge joint separation or interface tearing across the surface network. The obtained surface morphology corroborates the numerical outcomes demonstrated in [3] (Figure 7).

The geometric sensitivity of the surface framework is further checked by modifying the spatial positions of specific control points within the static mesh network. This operation evaluates the coordinate deformations and regional surface behavior under asymmetrical coordinate shifts to assess the localized shape control capacity. To analyze the vertex displacement patterns alongside the intensity distributions within a single visualization framework, the localized deformation analysis and the corresponding top-down color-mapped fields are generated as illustrated in Figure 8.

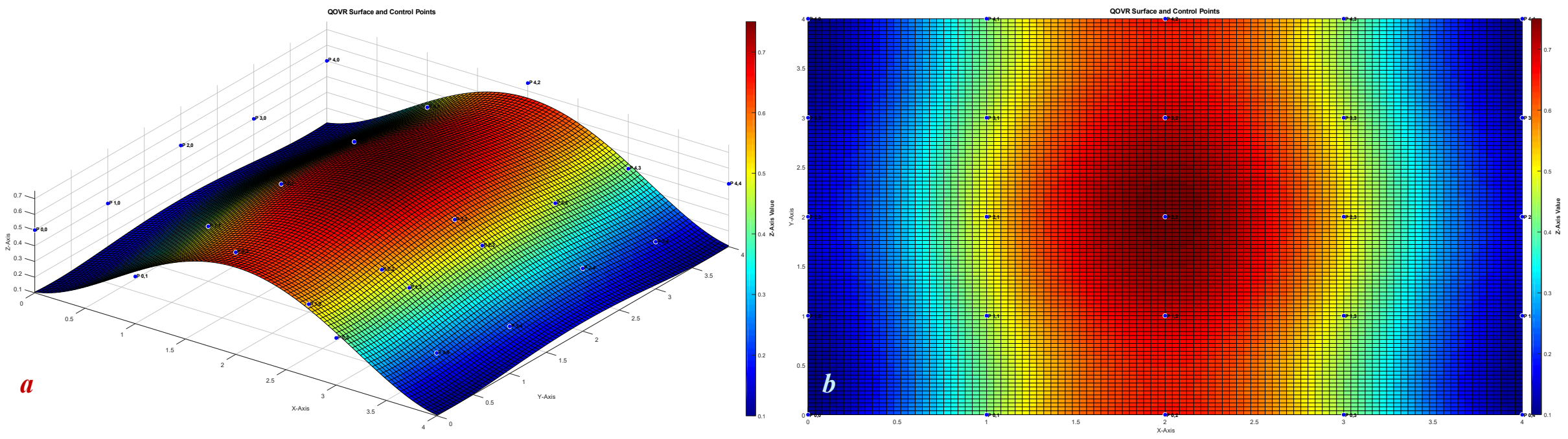

**Figure 8.** Parametric sensitivity and local shape adaptation variations under control point modification: (**a**) surface variation, (**b**) intensity distribution.

The physical and computational outputs observed in Figure 8 confirm the localized geometric modification capability under manual control point updates, demonstrating the utility of the surface framework within geometric environments. To evaluate the structural response, the vertical coordinate fields of the third column control points, $P_{0,2}, P_{1,2}, P_{2,2}, P_{3,2}$ and $P_{4,2}$, are modified by shifting their coordinates. The initial vertex values, starting at a uniform coordinate threshold of $Z = 0.50$ for the five network elements, are updated along a variable trajectory across the z-axis with heights shifting to $0.75, 0.15, 0.10, 0.15,$ and $0.75$. As shown in Figure 8a, this localized coordinate displacement constructs a parabolic configuration across the central surface segment, altering the surface framework.

The spatial concentration observed in the colormap visualization in Figure 8b verifies that the displacement effect decreases progressively as the distance from the perturbed control point increases. This behavior demonstrates the localized shape control capability of the surface framework, as the color-mapped intensity fields remain within the immediate neighborhood of the modified element. Consequently, the surrounding boundary regions maintain their initial structural continuity and spatial orientation without experiencing secondary curvature oscillations. This specific property aligns with the localized boundary conditions, showing that individual surface patches can be modified interactively while maintaining the localized boundary convergence of the entire network structure.

To evaluate the operational characteristics of the QOVR surface model, its parametric properties are compared against classical formulations, including B-spline, NURBS, Catmull–Rom, and Bézier approaches, as shown in Table 2. Traditional Bézier and B-spline surfaces operate under global connectivity rules, where modifying a single control point affects adjacent patch domains unless degree elevation or knot insertion routines are executed. While NURBS formulations provide shape control weights, they incorporate rational tracking layers at the node interfaces. Catmull–Rom surfaces permit localized interpolation directly through the node joints, maintaining fixed parameters at the boundaries without altering the internal knot vectors.

**Table 2.** Structural and operational comparison of surface formulation frameworks.

| Surface Formulation | Shape Control | Computational Basis | Boundary Behavior |
|---|---|---|---|
| Bézier/B-spline | Global | Polynomial degree elevation/Knot insertion | Prone to adjacent coordinate drift |
| NURBS | Weighted | Uniform rational tracking layers | Requires tracking at node interfaces |
| Catmull–Rom | Localized | Direct node joint interpolation | Lacks higher-order blending flexibility |
| Proposed QOVR | Localized | Symmetric matrix–tensor operations | Boundary position tracking via local parameters |

The QOVR framework incorporates direction-dependent parameter transformations directly into matrix–tensor operations. This architecture allows localized shape modifications along distinct grid lines while maintaining boundary alignment at the outer edge zones without requiring any global coordinate configuration adjustments. Computational outputs in Figure 8 confirm that the current surface framework isolates geometric deformations strictly within the targeted internal parameter configurations. This decoupling mechanism ensures that while local shape adaptation occurs internally, the boundary edges maintain positional convergence, which represents a characteristic of standard linear interpolation methods. The proposed QOVR isolates directional deformations via symmetric matrix–tensor operations to maintain structural symmetry across all boundary zones.

The baseline cubic Overhauser formulation utilizes a four-control point framework driving two local parabolas, as investigated in [4,8,9]. In comparison, the QOVR structure incorporates a five-control point network with three independent parabolas to form a baseline quartic matrix setup.

In summary, the geometric implementation and boundary analyses executed throughout Section 5 verify the operational capabilities of the QOVR surface framework. The sequential tracking workflow progresses from the initial mesh generation before and after transformation operations to the systematic evaluation of variable internal parameter thresholds. The outputs detailed from Figures 5–8 demonstrate that modifying the internal parametric knot vectors and shifting the control mesh coordinates yield localized shape variations across the target network zones. The outermost boundary tracks across all configuration sets maintain boundary alignment ensuring that surface closure is maintained. These graphical results and numerical tracking paths indicate that the localized blending tracks isolate coordinate deformations within the intended sectors, verifying the algorithmic framework within interactive geometric environments.

## 6. Conclusions and Future Work

This study presents the spatial and analytical construction of the QOVR surface generation framework designed to address boundary alignment and localized shape modification constraints. The QOVR surface framework introduces a variable-parameter fourth-degree architecture to achieve parameter tracking. Theoretical architecture tracks the single-parameter spline formulations, maps the orthogonal parameter transformations, and defines the surface mesh boundaries using symmetric matrix–tensor operations. The differential evaluations verify that the schema model satisfies zero-order position closure and first-order tangent vector continuity limits across both the exterior borders and the shared interior network junctions governed by the localized knot vectors. However, the algebraic derivations identify the structural boundaries of second-order curvature continuity at the mesh interfaces, establishing that higher-order smoothness is inherently unpreserved under static control point layouts due to the analytical mismatch between the left-hand and right-hand differential tracking paths.

The algorithmic processing steps are mapped onto a processing pipeline and evaluated using a graphical interface framework. The background software routine executes the surface coordinate generation based on a strict data dependency, where the raw surface grid data is computed prior to configuring the transformation matrix elements and exporting the transformed surface coordinate files. The graphical outputs demonstrate that the localized tracking layers allow independent geometric shape modification at the local network lines without triggering global polynomial degree elevation.

Beyond classical CAD/CAM applications, the structural flexibility and matrix-driven schema of the QOVR model offer functional deployment tracks for digital entertainment pipelines, PCG inside game design environments, and spatial visualization of multi-dimensional big data arrays. Future research lines will focus on extending the surface formulation to encompass arbitrary multi-sided topological patches and evaluating adaptive knot vector tracking algorithms to minimize geometric deformation limits under dynamic boundary constraints. Additionally, implementing parallel processing capabilities within the computational framework will be pursued to accelerate the transformation steps for real-time visualization of high-density geometric point clouds. In convergence with these computational scalability tracks, a prioritized numerical investigation focusing on formal error distributions, precise computational time benchmarks, numerical continuity tracking at node interfaces, and exact geometric deformation limits relative to conventional spline models under identical boundaries is also planned as a primary objective to complete the quantitative evaluation of the framework.

**Funding:** This research received no external funding.

**Data Availability Statement:** The GitHub repository contains the datasets and MATLAB (version R2018a) scripts used to implement the QOVR surface framework. This repository provides numerical data for the spatial control points matrix, untransformed raw surface network, boundary enclosure boxes, transformation matrix, and internal parametric knot vectors. These resources support the computational inspection and visualization of the exported coordinate tracking paths. https://github.com/HakanUst/QOVR-Surface. (accessed on 9 October 2026).

**Acknowledgments:** Generative AI (Google AI Search/Mode, driven by the Gemini model family) was used solely for English language editing, improving grammatical structure, and polishing the manuscript.

**Conflicts of Interest:** The author declares no conflicts of interest.